\documentclass{aa}  

\usepackage{graphicx}
\usepackage{txfonts}
\usepackage{amsmath}
\usepackage{natbib}
\usepackage{graphicx}
\usepackage{hyperref}
\usepackage{xcolor}
\usepackage{braket}
\usepackage{multirow}
\usepackage{subfig}
\hypersetup{
colorlinks=false, 
linkcolor=blue, 
urlcolor=red, 
linktoc=all 
}
\bibpunct{(}{)}{;}{a}{}{,}

\usepackage{comment}
\begin{document}


\title{Expanding Sgr~A* dynamical imaging capabilities with an African extension to the Event Horizon Telescope}
\titlerunning{Sgr~A* dynamical imaging with an African extension to the EHT}

\author{Noemi La Bella
          \inst{1}
          \and
          Sara Issaoun \inst{2,3,1} \and Freek Roelofs \inst{2,4,1} \and Christian Fromm \inst{5,6,7} \and  Heino Falcke \inst{1} 
          }

   \institute{Department of Astrophysics, Institute for Mathematics, Astrophysics and Particle Physics (IMAPP), Radboud University, P.O. Box 9010, 6500 GL Nijmegen, The Netherlands\\
   \email{n.labella@astro.ru.nl}
         \and
             Center for Astrophysics $|$ Harvard \& Smithsonian, 60 Garden Street, Cambridge, MA 02138, USA
        \and
        NASA Hubble Fellowship Program, Einstein Fellow
        \and
        Black Hole Initiative, Harvard University, 20 Garden Street, Cambridge, MA 02138, USA
        \and
        Institut f\"ur Theoretische Physik und Astrophysik, Universit\"at W\"urzburg, Emil-Fischer-Strasse 31, 97074 W\"urzburg, Germany
        \and
        Institut f\"ur Theoretische Physik, Goethe Universit\"at, Max-von-Laue-Str. 1, D-60438 Frankfurt, Germany 
        \and
        Max-Planck-Institut f\"ur Radioastronomie, Auf dem H\"ugel 69, D-53121 Bonn, Germany\\}
        

 
\abstract
{The Event Horizon Telescope (EHT) has recently published the first images of the supermassive black hole  at the center of our Galaxy, Sagittarius A* (Sgr~A*). Imaging Sgr~A* is plagued by two major challenges: variability on short (approximately minutes) timescales and interstellar scattering along our line of sight. While the scattering is well studied, the source variability continues to push the limits of current imaging algorithms. In particular, movie reconstructions are hindered by the sparse and time-variable coverage of the array.}
{In this paper, we study the impact of the planned Africa Millimetre Telescope (AMT, in Namibia) and Canary Islands telescope (CNI) additions to the time-dependent coverage and imaging fidelity of the EHT array. This African array addition to the EHT further increases the eastwest ($u,v$) coverage and provides a wider time window to perform high-fidelity movie reconstructions of Sgr~A*.}
{We generated synthetic observations of Sgr~A*'s accretion flow and used dynamical imaging techniques to create movie reconstructions of the source. To test the fidelity of our results, we used one general-relativistic magneto-hydrodynamic model of the accretion flow and jet to represent the quiescent state and one semi-analytic model of an orbiting hotspot to represent the flaring state.} {We found that the addition of the AMT alone offers a significant increase in the ($u,v$) coverage, leading to  robust averaged images during the first hours of the observating track. Moreover, we show that the combination of two telescopes on the African continent, in Namibia and in the Canary Islands, produces a very sensitive array to reconstruct the variability of Sgr~A* on horizon scales.}
{We conclude that the African expansion to the EHT increases the fidelity of high-resolution movie reconstructions of Sgr~A* to study gas dynamics near the event horizon.}

\maketitle
\section{Introduction} 

The Event Horizon Telescope (EHT) collaboration has recently published the first images of the black hole shadow of Sagittarius A* (Sgr~A*), the supermassive black hole (SMBH) at the center of the Milky Way, characterized by an asymmetric bright ring of ($52.1 \pm 0.6$) $\mu$as \citep{SgrA*PaperI}. The ring-like morphology was recovered in over 95$\%$ of the best-fit images produced from 2017 April 6 and 7 observations.
The EHT images of Sgr~A* are consistent with the prediction of a shadow for a Kerr black hole \citep{Falcke2000} with a mass $M \sim4 \times 10^6 M_\odot $ at a distance $ D \sim8$ kpc, which were accurately measured by high-resolution infrared studies of stellar orbits in the Galactic Center \citep{2018Gravity,Do2019}. In 2019, the EHT collaboration delivered the first ever image of a black hole shadow in the giant elliptical galaxy M87 \citep{M87paperI}. The main difference between the two SMBHs is their mass. M87* is about 1600 times more massive than Sgr~A*  and thus, it has a longer gravitational timescale. In fact, the period of the innermost stable circular orbit (ISCO) for a nonrotating black hole as massive as M87* is $\sim$30 days, while for Sgr~A* it is $\sim$30 minutes.
 As a consequence,  the estimation of the ring diameter of Sgr~A* is more uncertain than in M87* and we need movies to properly study the plasma motion surrounding the black hole on this short orbital timescale. 
 
The variability of Sgr~A* required a reformulation of the static source assumptions in the interferometric Earth aperture synthesis method and imaging algorithms used for M87*. In particular, to generate a typical static image of Sgr~A*, a variability noise budget needs to be added, while a dynamical imaging process is required to capture the evolving structure of Sgr~A* \citep{SgrAPaperIII}. Because of the sparsity of the EHT array, time slots with good ($u,v$) coverage were selected to perform dynamical studies on the variability \citep{2022Farah}.

The SMBH also presents flare events observed across the electromagnetic spectrum in the last decades. An accurate study of the millimeter light curves  during the 2017 EHT campaign was done by \cite{2022Wielgus}. In particular, the authors found excess variability on 2017 April 11, following a flare observed in the X-ray. Subsequent studies on polarized light curves with the Atacama Large Millimeter/submillimeter Array (ALMA) on the same day \citep{2022Wielgus2} revealed the presence of a hotspot orbiting Sgr~A* clockwise.

In addition to its quiescent variability, imaging Sgr~A* is a complex process because the very long baseline interferometry (VLBI) observations are affected by scattering in the interstellar medium along our line of sight toward the Galactic Center. The consequent diffractive and refractive effects of the scattering were mitigated by modeling their chromatic properties in the radio band \citep[see][for more details]{2018Psaltis,2018Johnson,Issaoun2019SgrA,Issaoun_2021,SgrAPaperIII}. 

Eight telescopes at six geographic locations formed the 2017 EHT array configuration that led to the first images of Sgr~A* and M87*. Since 2017, the array has doubled in bandwidth and increased the number of baselines with three new telescopes. 
As of 2022, the EHT has consisted of eleven telescopes at eight locations: ALMA and the Atacama Pathfinder Experiment (APEX) on the
Llano de Chajnantor in Chile; the Large Millimeter Telescope (LMT) Alfonso Serrano on the Volcán Sierra Negra in Mexico;
the James Clerk Maxwell Telescope (JCMT) and Submillimeter Array (SMA) on Maunakea in Hawai’i; the Institut
de Radioastronomie Millimétrique 30-m telescope on Pico Veleta (PV) in Spain; the Submillimeter Telescope (SMT) on Mt. Graham and the 12-m telescope on Kitt Peak (KP) in Arizona; the South Pole Telescope (SPT) in Antarctica; the Northern Extended Millimeter Array (NOEMA) in France; and the Greenland Telescope (GLT) at Thule.
This new configuration offers increased sensitivity of the array and will enable higher-fidelity images of Sgr~A* and M87*. However, all new telescopes are in the northern hemisphere and are less effective for imaging southern sources.
Additional telescopes are being considered to expand the capabilities of the array, especially on the African continent, which offers prime site locations to increase the ($u,v$) coverage toward Sgr~A*. 

In this work, we consider two additions to the EHT in the African region: one in Namibia and one in the Canary Islands. The Africa Millimetre Telescope (AMT),  planned on Mt. Gamsberg (2,347 m a.s.l.) in Namibia, will be the ﬁrst millimeter-wave telescope in Africa. The project to add this telescope is currently underway, and aims to relocate the decommissioned 15-meter SEST telescope in Chile to Gamsberg in the next years. This site will offer low precipitable water vapor levels during the typical fall and spring EHT campaign seasons \citep{2016Backes} and its strategical position in the southern hemisphere at the same latitude as ALMA provides important eastwest baselines to Chile and northsouth baselines to Europe, significantly increasing the snapshot coverage in the first half of a typical observing night. The island of La Palma in the Canary Islands (2,000 m a.s.l.) has dry weather conditions throughout the year \citep{2021Raymond} and offers a prime location to provide mid-range coverage between Namibia and Europe that is crucial to constrain source compactness and extent. Furthermore, the site's established infrastructure from existing observatories would make an additional telescope easily feasible and well supported, making it an ideal candidate for a telescope location in the near term. 

We present simulated dynamical images of Sgr~A* using the 2022EHT array and an African extension including two new telescopes: the 15-meter AMT, and the Canary Islands telescope, CNI, on the island of La Palma. We assume the dish size of CNI to be six meters, following the design concept for a next-generation EHT facility in the long term \citep{2019Doeleman}. We investigate the impact of the AMT and CNI stations on imaging Sgr~A* in both quiescent and flaring states. The methods we use can easily be expanded to other EHT configurations. 

The paper is organized as follows: in Section~\ref{Sec:Methods}, we describe the synthetic generation pipeline and imaging algorithms used. In Section~\ref{Sec:African array}, we present the African extension to the EHT and its contribution to snapshot and full-track ($u,v$) coverage. In Section~\ref{Sec:Imaging}, we show the static and dynamical reconstructions obtained with the enhanced EHT array. Finally, in Section~\ref{Sec:Conclusions}, we discuss the advantages of the African extension to the array in producing high-fidelity movies of Sgr~A*.

\section{Methods}
\label{Sec:Methods}

\subsection{GRMHD ground truth movies}
\label{Sec:models}
The quiescent state of the plasma flow of Sgr~A* was reproduced by generating synthetic data from general relativistic magneto-hydrodynamic (GRMHD) simulations at 230\,GHz. 
The typical range of simulations used to study Sgr A* include two classes of models: magnetically arrested disk \citep[MAD;][]{2003Igumenshchev, Narayan2003} and Standard And Normal Evolution \citep[SANE;][]{2012narayan} models. The SANE mode is characterized by a weak and turbulent magnetic field crossing the hemisphere of the event horizon, while the MAD mode has high magnetic flux. The recent EHT Sgr~A* results have shown that GRMHD simulations are more variable than the data \citep{SgrA*PaperV}. Because SANE models are less variable than MAD models, they are more representative of the degree of variability in Sgr~A*. We thus used a SANE model for our quiescent state reconstructions. 

The simulation was generated with the GRMHD code \texttt{BHAC} \cite{Porth2017, Olivares2020}. We initialized a torus in hydrodynamic equilibrium where the inner edge is located at 6\,M (where M is the gravitational timescale $GM/c^3$) and the pressure maximum is found at 13\,M. We set a black hole spin of $a_{\star}=0.9375$ and an adiabatic index $\hat\gamma=4/3$ and performed the simulations on spherical grid $(r,\theta, \phi)$ with resolution of $512\times192\times192$ and three layers of adaptive mesh refinement (AMR) using logarithmic Kerr-Schild coordinates. For more details on the simulations see \citet{Fromm2021}. We evolve the simulations until 30000\,M, which ensures a quasi steady-state in the mass accretion rate. The radiative transfer calculations were performed with the GRRT code \texttt{BHOSS}  \cite{Younsi2012,Younsi2016,Younsi2020,Younsi2021}. We used a field of view of 200\,$\rm{\mu as}$ together with a black hole mass of $4.14\times10^6\,M_\odot$ at a distance of 8.127\,kpc \citep{SgrA*PaperV}. The images were created assuming a viewing angle $\vartheta=10^\circ$ and a numerical resolution of 400$^2$ pixels. Since the electron temperature is not evolved during the GRMHD simulations we computed their temperature using the $R-\beta$ description of \citet{Moscibrodzka2016} where we set $R_{\rm low}=1$ and $R_{\rm high}=5$. In order to adjust the simulations to the observations, we iterated over the mass accretion rate to provide an average flux density of $\sim$2.4 Jy at 230\,GHz in a time window of 5000\,M. Two time windows were used (20-25\,kM and 25-30\,kM) and individually normalized.

The 16-hour movie consists of 300 frames separated by 200 seconds, with a rotation period of the plasma around the black hole of $\sim$30 minutes. The simulation does not include effects of interstellar scattering, therefore we characterized those effects using a phase screen toward Sgr~A*  \citep[see][for more details]{2018Psaltis,2018Johnson}. 

\subsection{Synthetic data generation}
\label{Sec:SYMBA}
The GRMHD synthetic data were produced with the SYMBA \footnote{
\url{https://bitbucket.org/M_Janssen/symba}} software \citep{Roelofs2020}, which reconstructs a model image following the same calibration and imaging processes of a realistic observation. Given a VLBI array configuration and a specific model as input, the synthetic observations are generated with MeqSilhouette \citep{Blecher2017, Natarajan2022} and the corrupted raw data are then processed with the VLBI data calibration pipeline rPICARD \citep{Janssen2019}, which is used to calibrate real EHT data \citep{EHT2019III}. The calibrated data set can also pass through the network calibration step that solves gains for colocated sites using the flux of the source at large scales \citep{2011Fish,2015Johnson&Gwinn,2019Blackburn,M87paperIII}.
Our synthetic data are based on the antenna and weather parameters as measured in the 2017 observations \citep{2019M87paperII}. The weather conditions were extracted from the VLBI monitor server \footnote{\url{https://bitbucket.org/vlbi}}, which collects weather data (e.g., ground pressure, ground temperature) from in situ measurements. 
The weather conditions used are reported in Table 2 of \cite{Roelofs2020}, which includes the parameters for the stations that joined the 2017 EHT campaign, and those for the enhanced array, with GLT joining the array in 2018, NOEMA and KP in 2021, and with the planned AMT. As described by the authors, the weather parameter estimation for stations that did not join the 2017 observations was done using the Modern-Era Retrospective Analysis for Research and Applications, version 2 ($\mathtt{MERRA-2}$) from the NASA Goddard Earth Sciences Data and Information Services Center \citep{Gelaro2017}, and the $\mathtt{am}$ atmospheric model software \cite{2019Paine}. We applied the same method to obtain the weather conditions on La Palma, in the Canary Islands. 
Finally, we adopted the observing schedule of 2017 April 7 \citep{SgrAPaperII}, encompassing scans on Sgr~A* from the 4 to 15 UT hours.  

For generating movies of flares in Sgr~A*, we used a simulated Gaussian flaring feature with an orbiting period of 27 min around a ray-traced image of a semi-analytic
advection-dominated accretion flow (ADAF) model of Sgr A* (model B of \citealt{2009Doeleman}). The movie at 230 GHz is composed of 100 frames separated by 16.2 seconds. The {\tt eht-imaging} Python library \footnote{\url{https://github.com/achael/eht-imaging}} \citep{Chael2016, Chael2018} was used to generate the hotspot synthetic data. The {\tt eht-imaging} package does not produce realistic VLBI-mm observations as SYMBA, for instance the data are not frequency-resolved, gain effects are not based on physical models, and there are no calibration effects added \citep[more details about the difference between the two pipelines can be found in][Appendix C]{M87paperIII}.
As in the case of the GRMHD movies, the synthetic data were based on the 2017 April 7 observing parameters. Unlike SYMBA, the simulated visibilities are not scan-separated, but have a cadence of 30 seconds.

\subsection{Dynamical imaging}
\label{Sec:Dynamicalimaging}
We imaged the SYMBA synthetic data set using the {\tt eht-imaging} library, developed specifically for the EHT. The imaging algorithm utilizes the regularized maximum likelihood (RML) method, which aims to find an image that minimizes a specified objective function, consisting of data fit quality ($\chi^2$) terms, and additional regularizer terms favoring, for example, smooth or sparse image structures \citep{2019M87paperIV}. The static assumption based on the Earth rotation aperture synthesis technique, where the source is assumed static during the course of the observation,  is not valid in the case of Sgr~A* due to its intraday variability \citep{SgrAPaperIII}.
To tackle this challenge, we use a method called ``dynamical imaging." The dynamical imaging algorithm within the {\tt eht-imaging} package is a generalization of the standard RML method which introduces three dynamical regularizers that enforce time-sensitive properties between snapshot frames \citep[see][for more details]{Johnson2017}. To reconstruct the hotspot movies we used the  $\mathcal{ R_{ \rm{\Delta} t}}$ regularizer, which imposes a time continuity between frames. Since the hotspot model simulates coherent motion of a flare orbiting Sgr~A*, this regularizer let us reconstruct continuous motion of structure. For the GRMHD movies, we also added the $\mathcal{ R_ {\rm{\Delta} I}}$ regularizer, which enforces similarity between the reconstructed frame and a time-averaged image. As GRMHD simulations describe the turbulent behavior of an accretion flow onto Sgr~A*,  this regularizer allows us to look for turbulence on top of a static structure. 

To inspect the capability of the expanded EHT array to reconstruct dynamical motion, we selected time windows during the observation for which coverage and filling fraction were optimized, as was done in \cite{2022Farah}.
For the GRMHD simulations, we produced movies of 5.7 hours, while for the hotspot movies we chose optimal time windows of 1.7 hours where the array offers the best coverage.  To obtain a good reconstructed movie, larger time windows were required for the GRMHD data set generated with SYMBA, which includes actual scans and gaps between the scans (more details in Section~\ref{Sec: uvcoverage}).

\subsection{Movie quality metrics}
\label{Sec:metrics}

Two quality metrics were selected to evaluate the fidelity of the reconstructed images: the normalized root-mean-square error (NRMSE) and the normalized cross-correlation \citep[NXCORR; e.g.,][]{2019M87paperIV}. NRMSE is more sensitive to pixel-by-pixel differences, while NXCORR is more sensitive to large scale structure \citep{Issaoun2019}. We estimated values for both metrics for each frame of the movie, quantifying the fidelity of the reconstruction as a function of time with respect to the ground truth. 

The NRMSE measures similarities per \textit{k}th pixel and it is defined as:
\begin{equation}
\begin{split}
\label{Eq:nrmse}
\rm{NRMSE} = \sqrt{\sum_{k} (I_{k} - I^{\prime}_{k})^2 \over \sum_{k} I_{k}^2},
\end{split}
\end{equation}
where I$^{\prime}$ and I are the intensity of the reconstructed movie frame and the model movie frame, respectively \citep[e.g.,][]{Chael2018, Issaoun2019}. An NRMSE value of zero corresponds to identical images. 

For given frames I$^{\prime}$ and I,  NXCORR is given by:
\begin{equation}
\label{Eq:nxcorr}
\begin{split}
{\rm{NXCORR}}  = {1 \over N} \sum_{k} \frac {(I_{k} - \Braket {I})(I^{\prime}_{k} - \Braket {I^{\prime}})} {\sigma_{I} \sigma_{I^{\prime}}},
\end{split}
\end{equation}
where \textit{N} is the total number of pixels per frame, $\Braket {I}$ and $\Braket {I^{\prime}}$ are the mean pixel values and $\sigma_{I},  \sigma_{I^{\prime}}$ are the  respective standard deviations. An NXCORR of 1 corresponds to a perfect correlation between the frames, -1 for anticorrelation, and 0 for no correlation \citep[e.g.,][]{2019M87paperIV}. 

\begin{figure}[t]
\centering
\resizebox{1. \hsize}{!}{\includegraphics{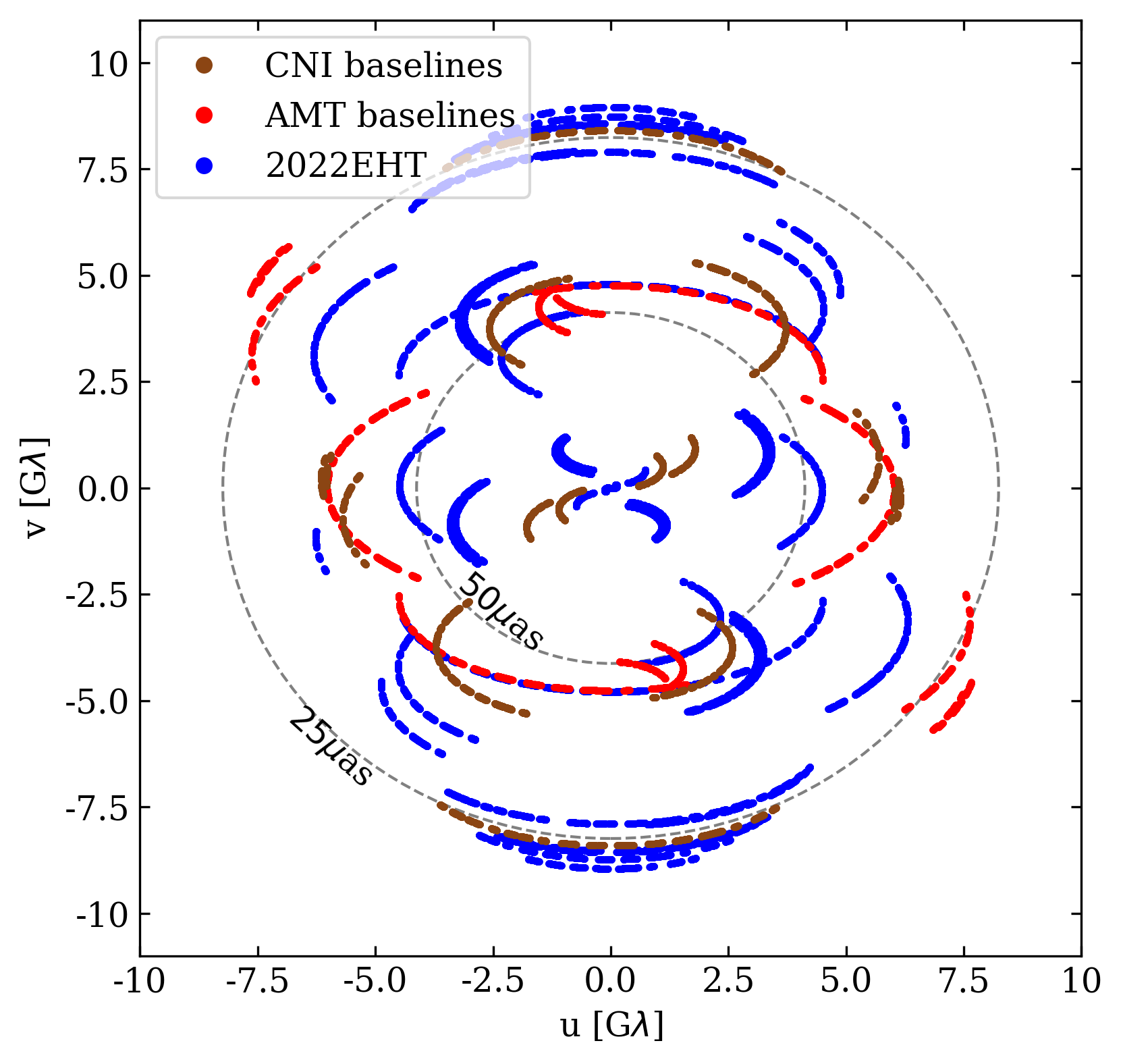}} \caption{Sgr~A* ($u,v$) coverage of the 2017 April 7 EHT observations. Seven scans on Sgr~A* were added to the original schedule at the beginning of the observation, brought by the introduction of the NOEMA array and the African stations. In blue, the coverage obtained with the 2022EHT array. The contributions of the AMT and CNI baselines are shown in red and in brown, respectively. The AMT adds long northeast and southwest baselines increasing the EHT resolution, while CNI offers shorter baselines to detect large-scale emission and constrain the source extent.}
  \label{fig:uvcoverage}
\end{figure}

\begin{figure*}[h]
   \centering
  \subfloat{\includegraphics[width=0.45\textwidth]{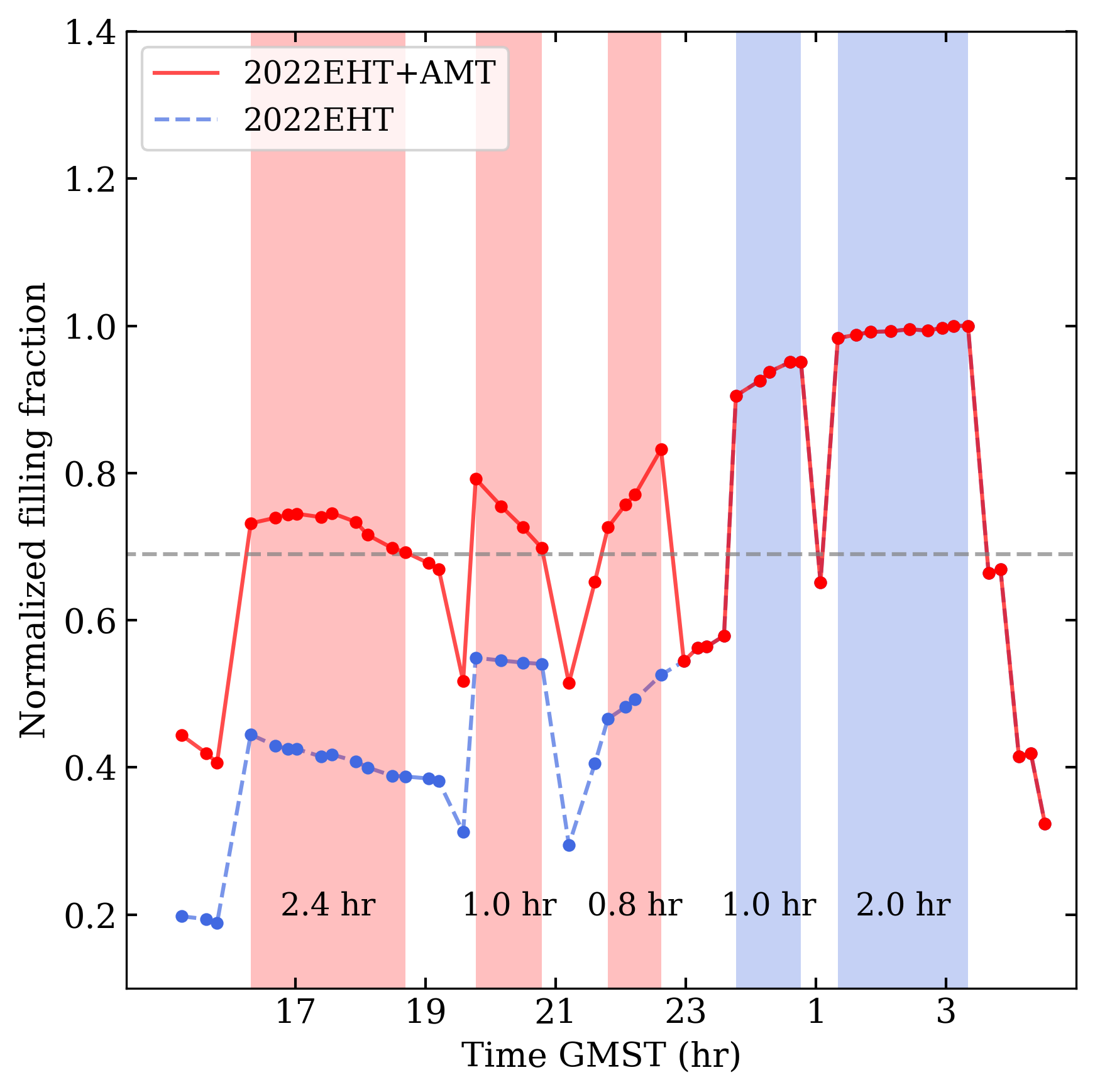}} \subfloat{\includegraphics[width=0.45\textwidth]{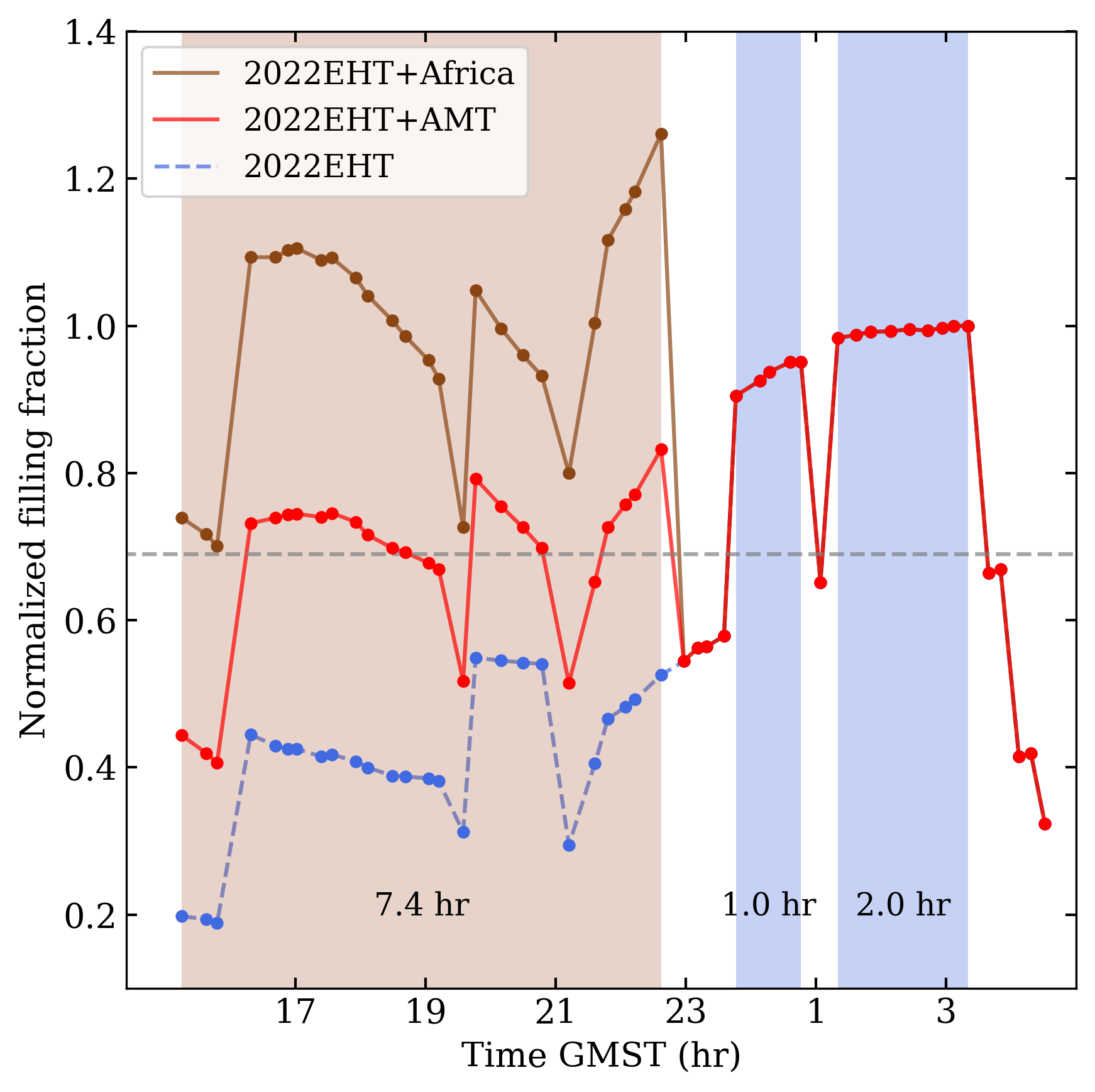}} 
   \caption{Time-dependent Fourier filling fraction normalized by the maximum Fourier filling of the 2022EHT array. The curves represent the filling fraction of the 2022EHT array, 2022EHT + AMT array and 2022EHT + Africa array, in blue,  red and brown, respectively. 
The dashed gray line corresponds to the lower limit used for identifying good time windows to perform dynamical imaging.
The optimal time regions for the current EHT array are shown in blue. The 2022EHT + AMT  adds three time windows (red areas) of $\sim$4 hours in total, while the 2022EHT + Africa array (brown area)  produces a time window of $\sim$7.4 hours.} \label{fig:filling_factor}
\end{figure*}

\begin{figure}[h]
   \centering
  \subfloat[]{\includegraphics[width=0.45\textwidth]{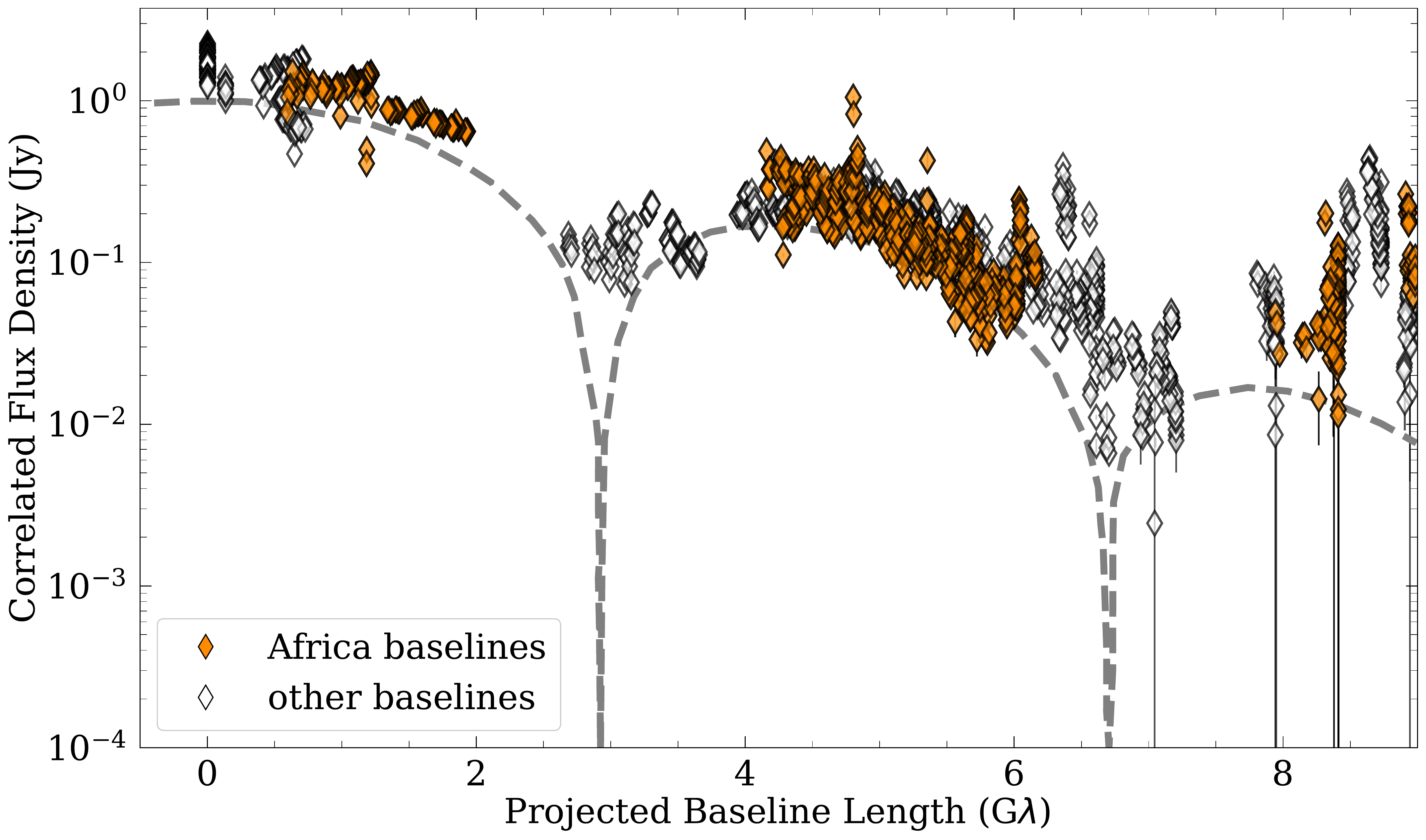}\label{fig:correlatedflux1}} \\  \subfloat[]{\includegraphics[width=0.45\textwidth]{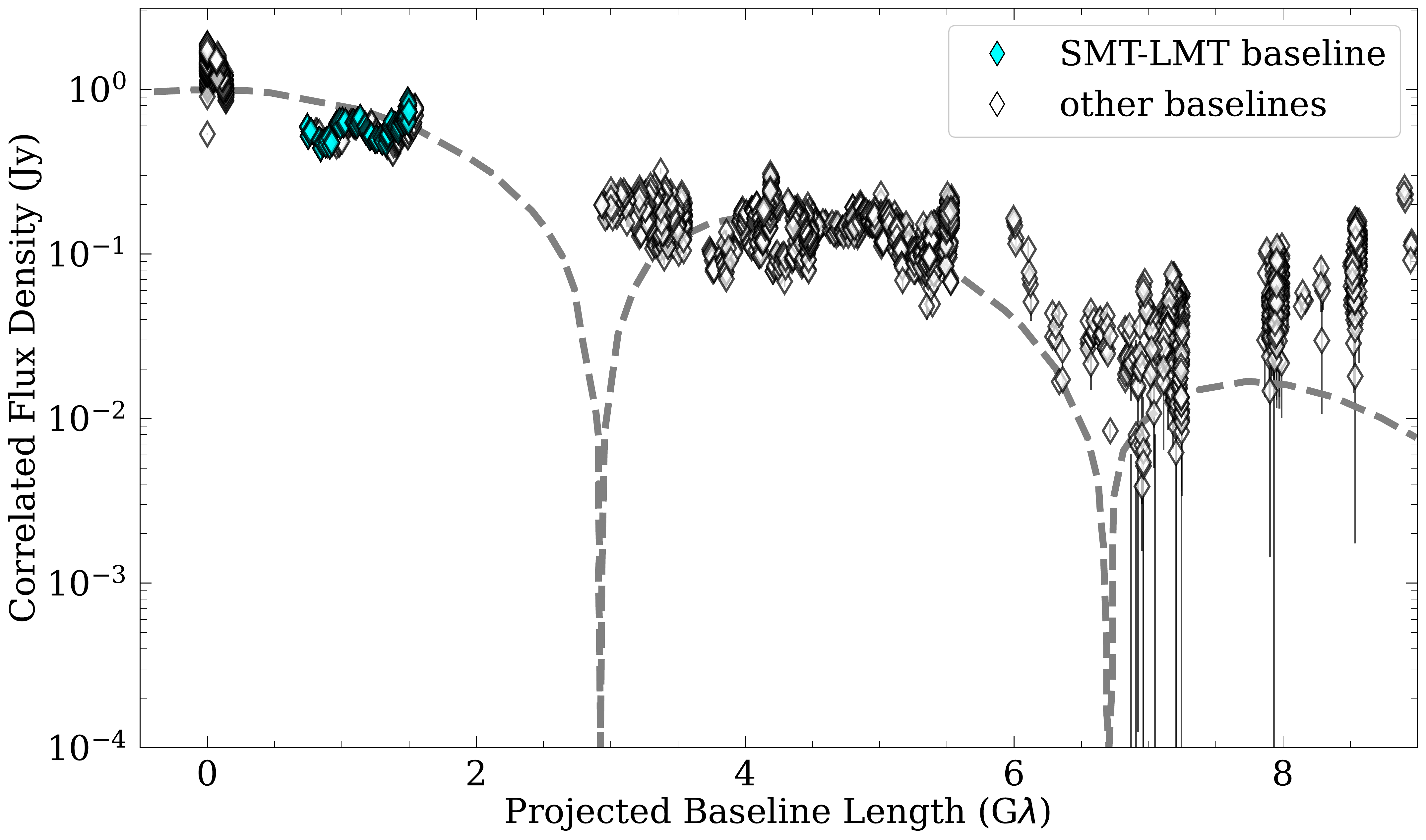}\label{fig:correlatedflux2}} 
   \caption{Correlated flux density as a function of baseline length for the Eastern (a) and Western (b) arrays. The African baselines (in orange) will contribute to probe the secondary peak, but also add short baselines to the array, at comparable projected baseline lengths to the SMT-LMT baseline (cyan). The shortest inter-site baselines are needed to estimate the extent and the total compact flux density of the source.}
\end{figure}

\section{The African expansion to the EHT}
\label{Sec:African array}

In this section, we discuss a potential implementation of the African expansion to the EHT, its ($u,v$) coverage, and Fourier filling fraction, which let us identify potential time windows to generate movies of Sgr~A*. We also investigate the location of the new baselines with respect to the position of the two local minima in the correlated flux density profile of a thin ring of 54 $\mu$as. To assess the impact of the new African stations, different array configurations were used. We name those configurations as follows: 2022EHT, the current EHT configuration composed of eleven telescopes; 2022EHT + AMT, the 2022EHT with the addition of AMT; 2022EHT + Africa, the 2022EHT plus the AMT and CNI stations; Eastern array + Africa, the 2022EHT subarray until $\sim$9.5 UT hours ($\sim$22.7 Greenwich Mean Sidereal Time, GMST), after this time the AMT does not observe Sgr~A*; Western array, the 2022EHT subarray from $\sim$9.5 UT hours to  $\sim$15 UT hours ($\sim$4.1 GMST). So far, the Western array has been offering the best coverage to produce dynamic reconstructions of Sgr~A*.

\subsection{($u,v$) coverage}
\label{Sec: uvcoverage}

Fig.~\ref{fig:uvcoverage} depicts the Sgr~A* ($u,v$) coverage using the 2017 April 7 observing schedule as a base, enhanced by the addition of NOEMA and KP, which joined the array post-2017, and the two proposed African antennas. Moreover, the observation was imposed to start when the source is at an elevation of more than 10 degrees at NOEMA and the African telescopes, allowing us to extend the observation by two hours. The 2022EHT baselines are shown in blue, the AMT baselines in red and the CNI baselines in brown. The AMT is a potential southern site to image Sgr~A* that adds determinant baselines to the array. Specifically, the AMT adds northsouth baselines to PV and NOEMA, eastwest baselines to Chile, and a redundancy baseline to ALMA-SPT, since Mt. Gamsberg is at the same latitude as ALMA.
Moreover, the AMT increases the resolution in the northeast and southwest, by adding long baselines to LMT and the Arizona stations. On the other hand, the CNI telescope yields new short inter-site baselines to the European sites, PV and NOEMA, contributing to the measurement of the source extent, together with the inter-sites SMT-LMT, PV-NOEMA baselines. In addition, the baseline CNI-AMT provides further northsouth coverage to the array. 

\subsection{Fourier filling fraction}
\label{Sec:fillingfactor}

The sparsity and changing coverage of the EHT array affect the accuracy of the dynamical reconstructions of time-variable sources. To produce VLBI movies of Sgr~A*, it is thus required to identify time periods with optimal and stable ($u,v$) coverage. For the 2017 Sgr~A* results, \citet{2022Farah} selected time regions using three different metrics and found the best dynamical time period to be from $\sim$01:30 GMST to $\sim$03:10 GMST, hence in the Western array window. We utilized one of these metrics, the ($u,v$) filling fraction, to inspect if new temporal regions are offered by the Eastern array + Africa. The Fourier filling fraction measures the area sampled in the ($u,v$) plane by the observed visibilities. Following \cite{2022Farah}, the ($u,v$) points were convolved with a circle of radius 0.71/$\theta_{\rm{FOV}}$,
with FOV being the field of view adopted for imaging,
representing the half-width at half-maximum of a filled disk of
uniform brightness on the sky \citep[see][for more details]{2019Palumbo}. In our analysis, we calculated the filling fraction normalized to the 
maximum filling fraction value of the 2022EHT array. On the left of Fig.~\ref{fig:filling_factor}, we show the time-dependent normalized filling fraction for the 2022EHT + AMT array in red, and that of the 2022EHT array in blue. The colored windows delimit time regions in which the filling fraction is persistently above the 70$\%$ 2022EHT maximum threshold (dashed grey line). Time windows below this threshold do not have sufficient coverage to produce high-fidelity movies. The 2022EHT array provides good time regions in the Western array. Notably, our results confirm the 01:30 GMST to 03:10 GMST best-time window obtained from the 2017 array selective dynamical imaging analysis \citep{2022Farah}. The AMT adds three additional optimal time periods  (red areas) in the Eastern array, of almost 4 hours in total. Furthermore, on the right of Fig. ~\ref{fig:filling_factor} we show a further increase in the Fourier filling area achieved by the combination of the CNI (brown) and AMT sites (i.e., with the 2022EHT array + Africa) leading to a persistent time block of 7.4 hours. Therefore, the African stations will provide significantly improved ($u,v$) coverage and stability for the Eastern array, increasing the ability to study rapid variations of the source at the beginning of the observing track. 

\label{Sec:static}
\subsection{Correlated flux density profile} 
The correlated flux density (in Jy) of Sgr~A* as a function of projected baseline length was investigated for both the Eastern and Western arrays using the network calibrated data sets obtained as output of SYMBA. The calibrated amplitudes of April 7, shown in Fig.\ref{fig:correlatedflux1} for the Eastern array and in Fig.\ref{fig:correlatedflux2} for the Western array, resemble a Bessel function with a first null at $\sim$3.0$\ G \rm{\lambda}$ and a second null at $\sim$6.5$ \ G \rm{\lambda}$, corresponding to a thin ring with a 54 $\mu$as diameter \citep{SgrAPaperIII}. In Fig.\ref{fig:correlatedflux1}, the African baselines, which are represented in orange, probe the prominent secondary peak. The African stations also provide short inter-site baselines at the same projected baseline length as the SMT-LMT baseline, highlighted in cyan in Fig.\ref{fig:correlatedflux2}. In 2017, the SMT-LMT baseline was the shortest inter-site baseline in the EHT array, which yields the size and the compact flux density estimation of the source \citep{Issaoun2019}. However, 2017 EHT observations have shown that LMT is a challenging station to calibrate and the determination of the compact flux is required to establish constraints on the data \citep{2019M87paperIV, SgrAPaperIII}. Since 2021, NOEMA and KP have added short baselines to PV and SMT, respectively, useful for amplitude calibration. Thus, the African baselines shorter than 2G$\lambda$ are important for the EHT imaging process as they can contribute to compute the size and the total compact flux density of the source. 

\begin{figure*}[h]
\centering
  \resizebox{1.\hsize}{!}{\includegraphics{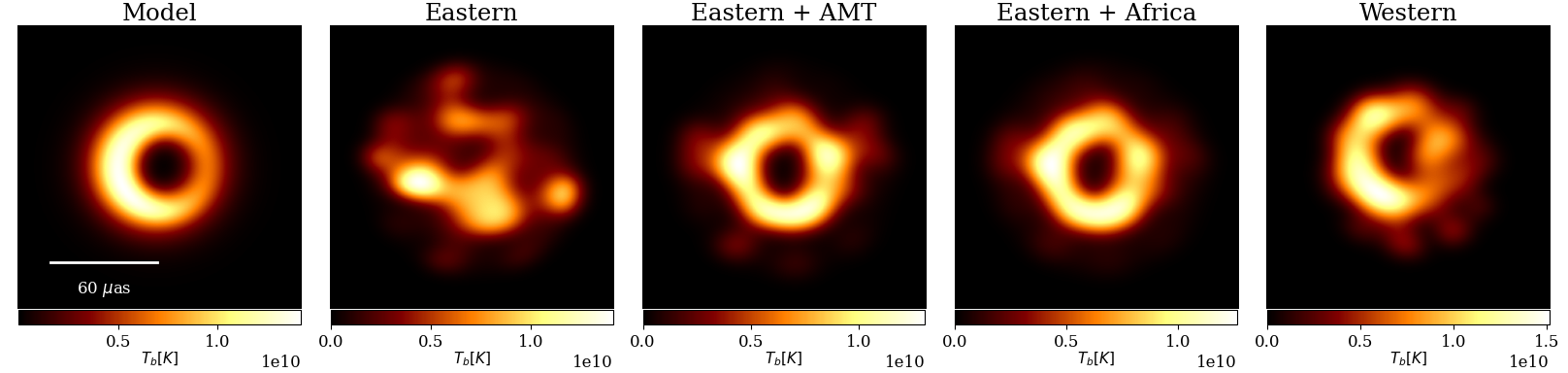}}
   \caption{Time-averaged reconstructions of Sgr~A* obtained from the GRMHD synthetic observations for the different array configurations. The leftmost image shows the static representation of the GRMHD simulation used as ground truth movie. The Eastern array without the AMT (second image) does not resolve the shadow of the black hole. The addition of the AMT significantly impacts the fidelity of the reconstruction, and a further improvement is obtained with the African array (third image). The rightmost image shows the averaged reconstruction produced using the Western array alone. The images were blurred with a Gaussian FWHM equal to 0.6 $\times$ clean beam of the 2022EHT + Africa data set. }
  \label{Fig:Static_images1}
\end{figure*}

\begin{figure} [h]
\centering
 \resizebox{1. \hsize}{!}{\includegraphics{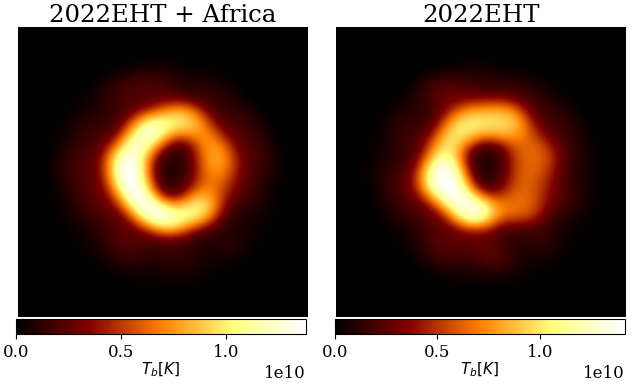}} \caption{Time-averaged reconstructions of GRMHD simulations of Sgr~A* with the 2022EHT + Africa and 2022EHT arrays. The ground truth model is shown in the first column of Fig.~\ref{Fig:Static_images1}. The 2022EHT + Africa array produces a higher fidelity image, which is used as the prior for the dynamical imaging. }
  \label{fig:static_2}
\end{figure}

\begin{figure*}[h]
\centering  \resizebox{1.\hsize}{!}{\includegraphics{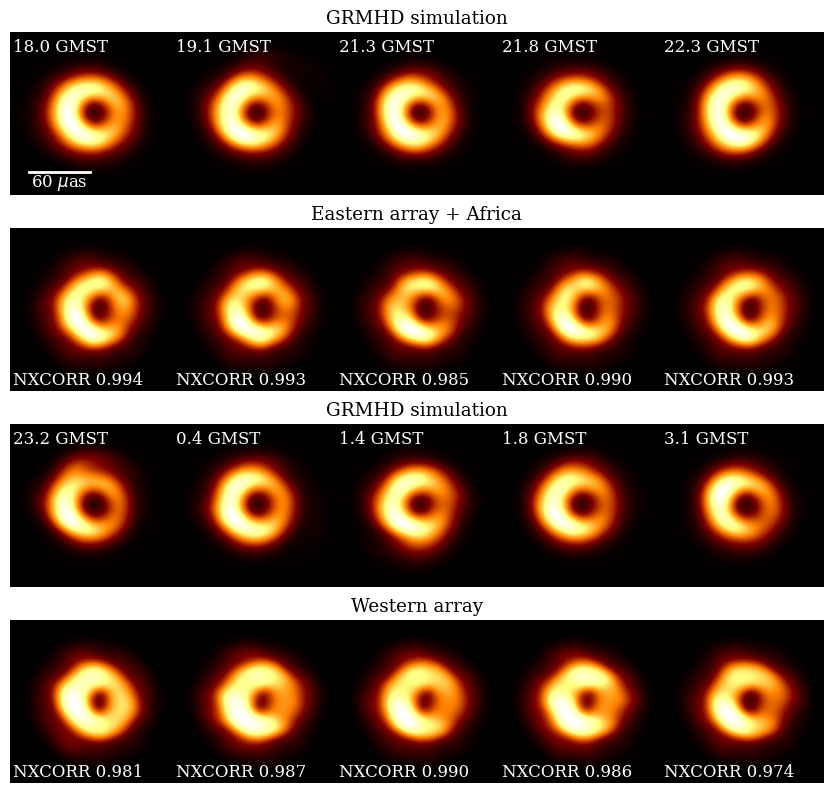}}
  \caption{Dynamical reconstructions obtained from the GRMHD data sets. The first row shows five snapshots of the GRMHD simulation taken in the Eastern array (17-22.7 GMST), the second row represents the respective dynamical reconstructions using the Eastern array + Africa. In a similar way, the third row and forth row illustrate the GRMHD frame simulations and the correspondent frame reconstructions using the Western array  (22.7-4.1 GMST). The blurring utilized for the GRMHD simulation is 0.6 $ \times$ clean beam. Higher quality dynamical reconstructions are produced by the Eastern array + Africa, also confirmed by the NXCORR metric reported at the bottom of each image. The numbers on the top of the GRMHD simulation snapshots represent the frame time.}
  \label{Fig:DI_SANE}
\end{figure*}

\begin{figure*}[h]
\centering
  \resizebox{1.\hsize}{!}{\includegraphics{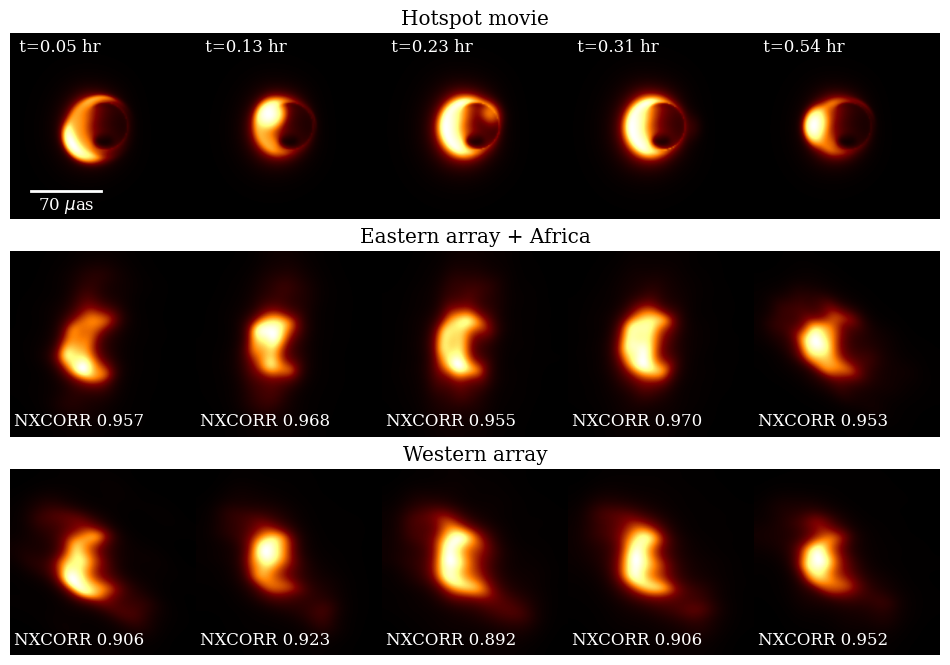}}
  \caption{Dynamical reconstructions generated using the hotspot synthetic data. In the first row we show five snapshots of the hotspot model movie. The hotspot performs a full rotation every 27 mins. The frames were chosen to represent a complete orbit. The reconstructions obtained from the dynamical imaging using the Eastern array + Africa and Western array are shown in the second and third row, respectively. The movies were generated in a time window of about 1.7 hours (21-22.7 GMST for the Eastern array, 1.5-3.2 GMST for the Western). The NXCORR values estimated for the reconstructions is reported in the bottom of each images. The temporal evolution is available as an online movie.}
  \label{Fig:DI_hotspot}
\end{figure*}

\section{Results from imaging}
\label{Sec:Imaging}
From the filling fraction study with the 2022EHT array + Africa, we estimated new time regions offered in the Eastern array to perform dynamical imaging. Here, we show the static and dynamical reconstructions from the GRMHD datasets generated with $\mathtt{SYMBA}$ using the Eastern array + Africa and Western array. Moreover, we present the dynamical reconstructions obtained from the hotspot model, which lets us test the capability of the array to image coherent motion or flares in Sgr~A*.
Unlike the ($u,v$) coverage inspection, the following images are obtained without the additional 2 hours observing Sgr~A* provided by the African stations. In this way, we compare the capabilities of the two subarrays to image Sgr~A* for the same observing time.

\subsection{GRMHD static reconstructions}
Fig.~\ref{Fig:Static_images1} shows the static images reconstructed from the GRMHD datasets for the different array configurations. The synthetic images were compared with the time-averaged image of the SANE simulation (first column), which was convolved with a Gaussian kernel with Full Width Half Maximum (FWHM) of 0.6 $\times$ clean  beam. 
As described in Sec.~\ref{Sec:Dynamicalimaging}, the static images were produced using the {\tt eht-imaging} package. We corrected for the effect of the diffractive scattering with the $\mathtt{eht-imaging \ deblur}$ function \citep{SgrAPaperIII}, which divides the interferometric visibilities by the Sgr~A* scattering kernel. 

Because the Eastern array without the African stations does not have sufficient coverage toward Sgr~A*, as we note from the filling fraction analysis,  it is not able to resolve its black hole shadow. The static reconstruction of Sgr~A* significantly improves when the AMT is added to the Eastern array, producing an image with a clear evidence of the ring-like structure. The image robustness increases with the Eastern array + Africa, indeed the artifacts present in the northwest and northeast of the ring are less evident than in the Eastern array + AMT image.
The averaged reconstruction using the Western array is also illustrated in the right-most side of the figure. The subarray is capable of
reconstructing the black hole shadow, but with a lower
quality than the Eastern array with the African stations. The fidelity of the reconstructions using the different array configurations well represents the filling fraction trend reported in Fig.~\ref{fig:filling_factor} and discussed in Sec.~\ref{Sec:fillingfactor}. 

In typical static imaging, the full observing track is used to produce the final averaged image. In Fig.~\ref{Fig:Static_images1}, we show segmented time-averaged reconstructions obtained with the Eastern and Western arrays individually with the purpose of examining the African station impact on imaging the static structure of Sgr~A*. The high-fidelity average image from the full 2022EHT + Africa array is illustrated on the left of Fig.~\ref{fig:static_2}, while on the right we show the static reconstruction using the 2022EHT array (see for comparison the representative model of Sgr~A* in Fig.~\ref{Fig:Static_images1}). 
The 2022EHT + Africa average image is used as the prior and initial image for the RML dynamical imaging of the GRMHD data sets presented in the next section.

\subsection{GRMHD dynamical reconstructions}
Movies of Sgr~A* were produced with the dynamical imaging algorithm introduced in Sec.~\ref{Sec:Methods}. 
Based on the candidate time regions with good ($u,v$) coverage explored in Sec.~\ref{Sec:Methods}, we produced movies for the Eastern and Western arrays, separately.
To perform dynamical imaging on the GRMHD data sets, which contain visibilities on a scan basis,  we chose large time periods of $\sim$5.7 hours, specifically from 17 GMST to 22.7 GMST for the Eastern array and from 22.7 GMST to 4.1 GMST for the Western array. The visibilities were averaged every 1 min to enhance the signal-to-noise ratio. 
The GRMHD simulation movie, which has a frame duration of 200 seconds, was synchronized to the reconstructed movies, which have a frame separation of 1 min. The synchronized model movie was created by averaging over the model frames that fall between the start and the end of the observed frame. In this way, we could estimate the NRMSE and NXCORR between the ground truth movie and the reconstructed movie frame by frame and select the data term and regularizer weights that minimize the NRMSE and maximize the NXCORR. 

In Fig.~\ref{Fig:DI_SANE}, we illustrate five snapshots of the movies reconstructed for the Eastern array + Africa (second row) and for the Western array (fourth row), and the corresponding frames of the SANE model. Each snapshot timestamp is shown at the top of the images. As for the static imaging, the reconstructions are descattered, by deblurring the interferometric data. The model movie was blurred using a Gaussian with a FWHM of 0.9 $\times$ clean beam of the 2022EHT + Africa data sets, while the reconstructions were blurred with a FWMH of 0.6 $\times$ clean beam. A lower blurring fraction is needed for the reconstructions because the dynamical imaging process inherently produces smoother structure.

The dynamical reconstructions generated with the Eastern array + Africa reproduce accurately the ring-like structure of the GRMHD simulation, while a less solid performance is obtained with the Western array. The reported NXCORR values in the bottom of the images confirm the robustness of the Eastern array + Africa reconstructions. The NRMSE values are also consistent with the general goodness trend of the reconstructions. 

We use GRMHD simulations of Sgr~A* to test if the Eastern array + Africa is able to reconstruct the main ring structure and its brightness distribution. GRMHD models reproduce a quiescent yet turbulent accretion flow and are not representative of coherent motion of features expected in the event of flaring activity. Moreover, GRMHD models are complex and challenging to reconstruct due to the large amplitudes in the variability \citep{SgrA*PaperV}, making it difficult to recognize the rotation of individual features. Dynamical imaging using a simple hotspot model, shown in the next section, allows us to easily investigate the capability of the array to reconstruct coherent motion in Sgr~A* in the event of flares.

\begin{figure*}[h]
\centering
  \resizebox{1.\hsize}{!}{\includegraphics{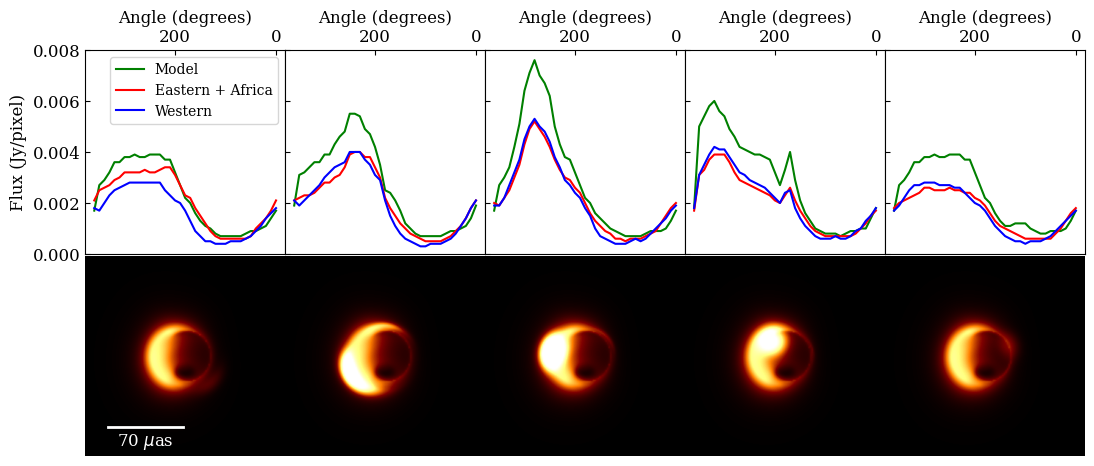}}
  \caption{Flux density (Jy/pixel) in function of the angle (degrees) estimated in five snapshots of the model movie (in green), of the Eastern array + Africa movie (in red), and of the Western array movie (in blue). The brightness distribution was estimated using a ring with outer radius of 32 $\mu$as, divided in sectors 10 degrees wide. The five frames of the model simulation from where the flux densities were extracted are shown in the bottom panel.}
  \label{Fig:flux_profiles}
\end{figure*}

\begin{figure*}[h]
   \centering
  \subfloat{\includegraphics[width=0.5\textwidth]{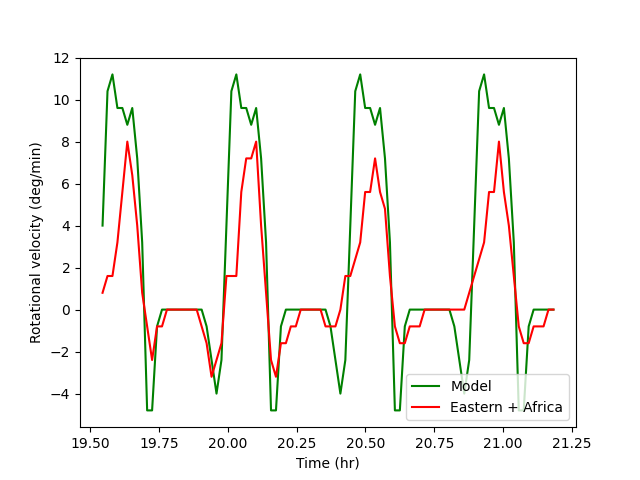}} \subfloat{\includegraphics[width=0.5\textwidth]{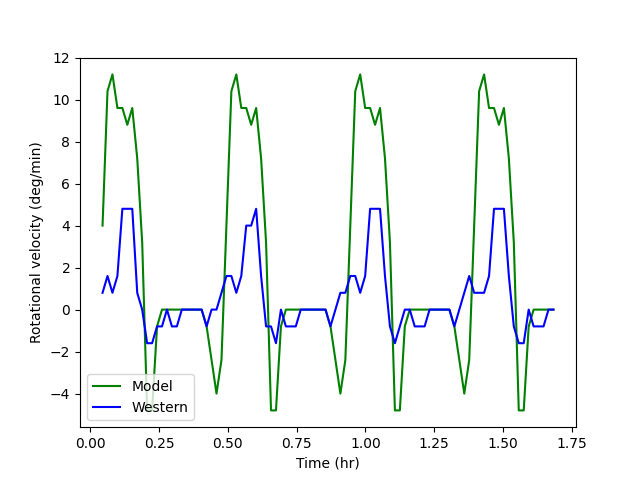}} 
   \caption{Rotational velocity (degree per minute) in function of the time for the Eastern + Africa array movie (left) and for the Western array movie (right). In green, the rotational velocity for the hotspot movie simulation. The profile were obtained by searching for the angle that maximize the similarity between each frame and the subsequent fifth frame. The Eastern array + Africa movie presents a more robust reconstruction of the hotspot rotation than the Western array. The negative values of the rotation are artifacts produced by the method utilized.} 
\label{Fig:rotationalvelocitu}
\end{figure*}

\subsection{Hotspot dynamical reconstructions}
Fig.~\ref{Fig:DI_hotspot} shows five snapshots of the dynamical reconstructions generated using as ground truth the hotspot crescent model. In the first row, we present the synchronized model snapshots, while the Eastern array + Africa and Western array dynamical reconstructions are illustrated in the second and third row, respectively. Similarly to the GRMHD models, we identified the data terms and regularizer weights that maximize the similarities between the model and the reconstruction snapshots, exploiting the NXCORR and NRMSE metrics.
Unlike the GRMHD reconstructions, the visibilities are separated by $\sim$30 seconds and the dynamical imaging was performed in narrow time regions of about 1.7 hours. In particular, for the Eastern array + Africa this was chosen to be from 21 to 22.7 GMST, which corresponds to the best time window offered by the subarray. For the Western array, the best period is given between the 1.5 and 3.2 GMST. The five snapshots are separated by almost 0.1 hour in order to represent the hotspot orbit, which is completed in $\sim$0.5 hours (i.e., 27 minutes). As confirmed by the NXCORR (reported in the figure) and the NRMSE, the individual frames produced in the Eastern array + Africa time region are more accurate than in the Western array window. Indeed, in the latter, the snapshots present pronounced northeast and southwest imaging artifacts.
Both subarrays are capable of reconstructing the motion of the hotspot, confirming that the addition of the African stations to the EHT array provides a new time window in the first half of the observation to detect rapid coherent flux variations in Sgr~A*'s accretion flow or jet. In order to effectively establish the capability of the array in reproducing the flare motion, we developed two methods that evaluate the robustness of our dynamical images. For the two subarrays, we investigate the ability to recover the flux density profile and the time-dependent rotational velocity of the hotspot. The two methods are described in Sec.~\ref{Sec:fluxprofile} and in Sec.~\ref{Sec:rotationprofile}.

\subsubsection{Method 1: Flux density profile}
\label{Sec:fluxprofile}
To assess the ability of the Eastern array + Africa to reconstruct the flux density around the crescent model, we calculated the flux density pixel by pixel as a function of the position angle for each snapshot. We selected the ring from which to extract the flux using the $\mathtt{hough\_ring}$ function in the {\tt eht-imaging} library, which finds circles in an image according to the pixel brightness distribution. The choice was made giving as input the time-averaged model image. Thus, for each model snapshots and reconstructed frames, the flux density was estimated within a radius of 32$\mu$as and in sectors 10 degrees wide. In Fig.~\ref{Fig:flux_profiles}, we show the flux density as a function of the angle for five snapshots of the ground truth model (in green and also illustrated in the lower panel of the image), of the Eastern array + Africa movie (in red) and of the Western array movie (in blue).
Because of the asymmetry of the brightness distribution in the crescent model, the flux profile has a peak in the snapshots when the hotspot is at its maximum intensity (i.e., third column), while it decreases when the hotspot is located on the opposite side. From the model snapshots and the corresponding flux density profile, we note that the angular position of the hotspot is correctly determined by this method. The flux density profiles obtained with the Eastern array + Africa and Western array recover quite well the hotspot motion, both in term of intensity and in position angle.

\subsubsection{Method 2: Rotational velocity profile}
\label{Sec:rotationprofile}
Additionally, we computed the rotational velocity of the hotspot as a function of time. This rotation (in degrees per minute) is defined as the degree of rotation for each frame $i$ with respect to the fifth subsequent frame $j$. In order to measure it, we rotated frame $i$ in steps of two degrees across a range of angles. We calculated the NXCORR (i.e., the image correspondence) with respect to frame $j$ at each rotation angle. The angle at which the NXCORR is maximized between the two frames divided by the time duration between frames $i$ and $j$ gives us the rotational velocity. We measured the rotational velocity of the hotspot every five frames, which lets us reconstruct its motion. As the hotspot completes its orbit every 27 min and the frame separation of the reconstructed movie is $\sim$30 seconds, the rotation every five frames ($\sim$33$^\circ$) is easier to measure than the rotation per frame ($\sim$6.6$^\circ$).

The rotational velocity obtained for the Eastern array + Africa and the Western array movies are shown in the left and right of Figure~\ref{Fig:rotationalvelocitu} in red and in blue, respectively. The hotspot velocity measured from the model movie is represented in green. As in the case of the flux profile, the method represents the asymmetric brightness distribution of the crescent model. Indeed, the frames with the maximum intensity of the hotspot have a maximum value of the rotational velocity, which drops to zero when the hotspot is not present. The negative values of the velocity are artifact produced by the method. In particular, these unphysical features are generated for each period of the hotspot movie, when we compare the last frame that contains the hotspot and the fifth frame that presents only the crescent emission. Comparing the rotational velocity curves derived from the Eastern array + Africa and the Western array movies with the model simulation, we find that the flare variability is most accurately recovered in the Eastern time window.

\section{Summary and conclusions}
We generated synthetic data of Sgr~A* with the current EHT array and two stations in the African continent, the AMT and the CNI telescope. We have evaluated the capability of the EHT Eastern subarray with the African sites (17-22.7 GMST) to produce movies of Sgr~A* and compared it to the Western subarray (22.7-4.1 GMST). The data sets were created from ray-traced images of a SANE GRMHD simulation, which is representative of the quiescent yet turbulent black hole accretion flow, and from a crescent hotspot model to test the imaging performance of the array in reconstructing coherent motion from flaring activity in Sgr~A*. \\

We found that the AMT increases the resolution of the EHT array via long baselines with the Arizona and Mexico sites, while short baselines provided by the African extension to the EHT constrain the compactness and extent of the source on larger scales. We estimated  the Fourier filling fraction with the EHT array and the Africa telescopes to investigate the presence of good time regions to perform dynamical imaging. We found that the added baselines offer an optimal time window of about 7 hours in the Eastern array, allowing to produce high-fidelity movies of Sgr~A* from the very start of a typical observing track. This increases the time in which dynamical imaging is possible by a factor > 4. In comparison, \citet{2022Farah} demonstrated that with the 2017 EHT array, the only time period in which we are able to reconstruct the variability of the source is from $\sim$01:30 GMST to $\sim$03:10 GMST, with the Western array.

Our static reconstructions of the GRMHD simulation confirm the importance of the AMT in imaging Sgr~A*. Without the AMT, the data set generated with the current EHT configuration is not able to reproduce a physical image of the black hole shadow in the Eastern array window. Including the African sites, we can perform high-fidelity imaging of Sgr~A* with reduced artifacts. Additionally, we produced GRMHD dynamical reconstructions limited to the best Eastern and Western time regions. The African stations enable accurate frame reconstructions of the ring-like structure when included in the Eastern array.
Since the rotation of individual features is difficult be recognized in the turbulent flow of GRMHD simulations, we performed a hotspot dynamical imaging analysis to test the capability of the different arrays to reconstruct coherent motion mimicking flaring activity in Sgr~A*. Compared to the 2022EHT array, the African stations open a new time window in the Eastern array that can be used to reconstruct motion in the accretion disk. We developed two methods involving the determination of the flux density profile and the rotational velocity of the hotspot to establish the successful performance of the enhanced Eastern array in reproducing the motion in Sgr~A*. Our results show the impact of adding stations in the African continent in increasing the time-variable ($u,v$) coverage of the EHT toward Sgr~A*. The African extension will be crucial for future EHT observations to study accurately the time-variable source at our Galactic Center through high-fidelity movies across an observing track.

\label{Sec:Conclusions}

\begin{acknowledgements}
We thank Oliver Porth for performing the ray-tracing for the GRMHD simulation used. This publication is part of the project Dutch Black Hole Consortium (with project number 1292.19.202) of the research programme NWA which is (partly) financed by the Dutch Research Council (NWO). SI is supported by Hubble Fellowship grant HST-HF2-51482.001-A awarded by the Space Telescope Science Institute, which is operated by the Association of Universities for Research in Astronomy, Inc., for NASA, under contract NAS5-26555. FR is supported by NSF grants AST-1935980 and AST-2034306, and the Black Hole Initiative at Harvard University, made possible through the support of grants from the Gordon and Betty Moore Foundation and the John Templeton Foundation. The opinions expressed in this publication are those of the author(s) and do not necessarily reflect the views of the Moore or Templeton Foundations. CMF is supported by the DFG research grant ``Jet physics on horizon scales and beyond" (Grant No.  FR 4069/2-1) The simulations were performed on LOEWE at the CSC-Frankfurt, Iboga at ITP Frankfurt and Pi2.0 at Shanghai Jiao Tong University. 

\end{acknowledgements}

\bibliographystyle{aa.bst} 
\bibliography{biblio.bib} 

\end{document}